\documentclass{article}

\PassOptionsToPackage{numbers,compress}{natbib}
 \usepackage[preprint]{neurips_2026}

\usepackage[utf8]{inputenc} % allow utf-8 input
\usepackage[T1]{fontenc}    % use 8-bit T1 fonts
\usepackage{hyperref}       % hyperlinks
\usepackage{url}            % simple URL typesetting
\usepackage{booktabs}       % professional-quality tables
\usepackage{amsfonts}       % blackboard math symbols
\usepackage{nicefrac}       % compact symbols for 1/2, etc.
\usepackage{microtype}      % microtypography
\usepackage{xcolor}         % colors
\usepackage{amsmath}
\usepackage{amsmath,amssymb,amsfonts}
\usepackage{algorithm}
\usepackage{algpseudocode}
\usepackage{graphicx}
\usepackage{caption}
\usepackage{enumitem}
 \usepackage{array}
 \usepackage{placeins}
\usepackage{cuted}
\usepackage{float}
\usepackage{makecell}
\usepackage{tabularx}

\title{FATE: Future-State-Aware Scheduling for Heterogeneous LLM Workflows}

\author{%
  Zirui Huang, Yi-Xiang Hu, Feng Wu, Xiang-Yang Li \\
  School of Computer Science and Technology\\
  University of Science and Technology of China, Hefei, China\\
  \texttt{\{hzr090901,yixianghu\}@mail.ustc.edu.cn,\{wufeng02,xiangyangli\}@ustc.edu.cn}
}

\begin{document}

\maketitle

\begin{abstract}
Large language model (LLM) applications are increasingly executed as heterogeneous multi-stage workflows rather than isolated inference calls. In these workflow directed acyclic graphs (DAGs), scheduling decisions affect not only the currently ready stage, but also the execution state inherited by downstream stages, including model residency, parent-output locality, prefix reuse, and future device reachability. Existing serving and DAG-scheduling policies mainly optimize immediate queue state, placement cost, or reuse signals in isolation, which can fragment useful state and increase end-to-end latency. We present \textsc{FATE}, a future-state-aware scheduler for heterogeneous LLM workflows. \textsc{FATE} combines a CP-SAT-backed frontier planner, horizon-aware candidate scoring, bounded multi-device shard execution, and state-conditional cost estimation. Rather than solving a monolithic full-DAG problem, \textsc{FATE} repeatedly plans over the current ready frontier and scores assignments by both immediate cost and the downstream state they induce. Across a WfCommons-derived workflow-DAG benchmark and controlled prefix-reuse benchmarks, \textsc{FATE} outperforms practical heuristics, classical DAG scheduling, and proxy adaptations of recent workflow-serving policies. On the workflow-DAG benchmark, it achieves normalized makespan and normalized P95 latency of 0.675 and 0.677, reducing them by 32.5\% and 32.3\% over RoundRobin and by 8.9\% and 8.8\% over the strongest non-\textsc{FATE} baseline. Mechanism analysis and ablations show that these gains arise from jointly preserving multiple dimensions of future execution state rather than prefix reuse alone. These results suggest that future-state preservation is a useful scheduling objective for heterogeneous LLM workflow serving.
\end{abstract}

\section{Introduction}
Recent advances in large language model (LLM) serving have substantially improved the efficiency of individual inference calls. Systems for paged KV-cache management, structured language-program execution, prefill/decode scheduling, and architectural disaggregation have improved throughput, latency, and memory efficiency~\cite{kwon2023efficient,zheng2024sglang,agrawal2024taming,zhong2024distserve}. These techniques are highly effective when a workload is well approximated as a single model invocation or a weakly coupled sequence of calls.

Modern LLM applications increasingly break this abstraction. A single user query can trigger a dependent workflow involving retrieval, decomposition, routing, tool use, code execution, verification, aggregation, and post-processing. This pattern is especially visible in agentic and software-engineering assistants, where a high-level request may be decomposed into specialized planning, implementation, testing, review, and repair stages. These applications are naturally represented as directed acyclic graphs (DAGs) with heterogeneous stages, heterogeneous models, and explicit cross-stage dependencies. Recent systems have started to expose this structure directly, treating agentic applications as workflow-serving problems rather than as independent requests~\cite{luo2025autellix,chaudhry2025murakkab,laju2026nalar}.

This shift changes the scheduling problem fundamentally. In conventional serving, a scheduler mainly reacts to queue state: which requests are ready, which device is idle, and how batching should be arranged. In heterogeneous workflow serving, by contrast, the scheduler also shapes the execution state inherited by downstream stages. Placing a stage on one device rather than another may preserve model residency, avoid a future transfer, improve parent-child colocation, or retain reusable prefix-related state for downstream stages~\cite{pankvflow,mei2025helix}. Conversely, aggressively balancing current load can destroy locality that matters more for end-to-end workflow latency.

\begin{figure}[t]
    \centering
    \includegraphics[width=1\linewidth]{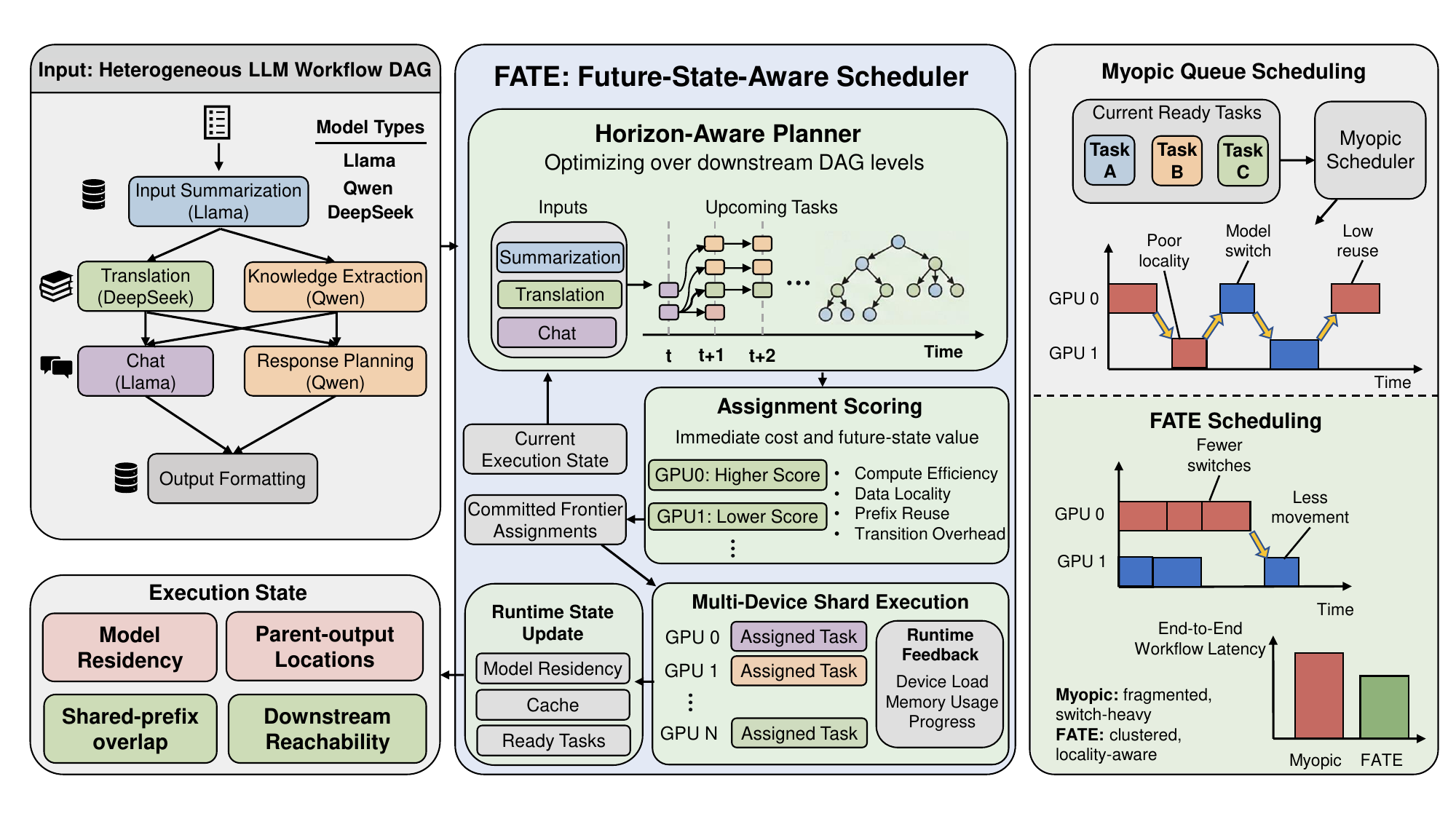}
    \caption{Overview of \textsc{FATE}. Given a heterogeneous workflow DAG and the current execution state, \textsc{FATE} performs horizon-aware planning, scores feasible stage-device assignments, executes bounded multi-device shard placements, and updates execution state as execution unfolds. Unlike myopic queue-only scheduling, \textsc{FATE} optimizes both immediate cost and future-state quality.}
    \label{fig:fate_overview}
\end{figure}

We argue that heterogeneous LLM workflow scheduling should therefore optimize not only immediate execution efficiency, but also \emph{future-state preservation}. We call the resulting scheduler \textsc{FATE}, short for \emph{Future-stATE-aware} scheduling. Figure~\ref{fig:fate_overview} provides an overview. Given a heterogeneous workflow DAG and the current execution state, \textsc{FATE} repeatedly chooses assignments for the current ready frontier while scoring each candidate by the future execution state that it would create: which model remains resident, which parent outputs stay local, which prefixes remain aligned with downstream stages, and which later stages remain reachable without avoidable transfer. A limited downstream horizon estimates these consequences, but the planner commits only the current frontier before advancing and updating state. Unlike purely greedy or rank-based DAG schedulers such as HEFT~\cite{topcuoglu2002performance}, \textsc{FATE} explicitly treats state-dependent downstream cost as part of the scheduling objective~\cite{luo2025autellix,chaudhry2025murakkab,pankvflow}.

Empirically, we find two robust results. First, execution-state-aware scheduling improves end-to-end workflow performance across our WfCommons-derived workflow-DAG benchmark~\cite{coleman2022wfcommons}. Second, these gains are not explained by prefix reuse alone: in the controlled prefix-reuse suite, reuse-aware baselines narrow the gap but do not close it. Together, these results suggest that future-state preservation is a useful scheduling objective for the workflow-serving regimes we study.

Our main contributions are as follows:
\begin{itemize}[leftmargin=*, labelsep=0.5em]
    \item We identify \emph{execution state} as  a useful and often under-optimized objective for heterogeneous LLM workflows. Unlike conventional serving, where scheduling mainly reacts to queue state, workflow execution depends on evolving model residency, cross-stage locality, reusable prefix-related state, and constrained future reachability.
    \item We formalize execution-state-aware placement and scoring for heterogeneous workflow DAGs, capturing stage placement, bounded multi-device execution, switching costs, transfer costs, and state-dependent execution benefits under device constraints.
    \item We instantiate this formulation with a CP-SAT-backed~\cite{perron2023cp} scheduler with horizon-aware planning and state-conditional scoring, and show that the central tradeoff is not simply better balancing, but balancing versus future-state preservation.
    \item We introduce a unified heterogeneous multi-LLM workflow scheduling framework that makes these state variables explicit and allows future-state-aware policies, practical heuristics, classical DAG schedulers, and proxy adaptations of recent serving policies to be evaluated under the same workflow runtime.
\end{itemize}

\section{Problem Formulation}
\label{sec:formulation}

We consider an online system serving a dynamic set of workflow instances. Let $\mathcal{W}_t$ denote the set of active workflows at time $t$. For clarity, we define stage- and workflow-level quantities for a generic workflow instance and suppress workflow indices when unambiguous. Runtime ready sets, frontier sets, and feasible action sets are therefore understood as the unions of the corresponding per-workflow sets over the currently active instances in $\mathcal{W}_t$. Each workflow instance is represented as a DAG $G=(V,E)$, where each node $v\in V$ denotes a workflow stage and each directed edge $(u,v)\in E$ denotes a precedence dependency, i.e., stage $v$ can start only after receiving the output of stage $u$.

For each stage $v$, let
\[
\big(m(v),\, A(v),\, R(v),\, c_v,\, \phi(v),\, \mathrm{Pa}(v),\, \mathrm{Ch}(v)\big)
\]
denote its attributes, where:
(i) $m(v)$ is the model type used by $v$;
(ii) $A(v)\subseteq \mathcal{D}$ is the set of eligible devices;
(iii) $R(v)\in \mathbb{Z}_{\ge 1}$ is the maximum degree of bounded multi-device shard execution;
(iv) $c_v:A(v)\rightarrow \mathbb{R}_{+}$ is the base runtime profile across eligible devices, with
\[
c_{\text{base}}(v,d)=c_v(d),\qquad d\in A(v);
\]
(v) $\phi(v)$ denotes stage-local features, including prompt metadata, shared-prefix-group information, cache-related flags, and runtime-side stage annotations;
and (vi) $\mathrm{Pa}(v)$ and $\mathrm{Ch}(v)$ denote the parent and child stages of $v$.

At time $t$, the execution state is denoted by
\[
s_t = (\rho_t,\kappa_t,\ell_t,\tau_t),
\]
where $\rho_t$ encodes model residency on each device, $\kappa_t$ encodes reusable prefix-related metadata on each device, $\ell_t$ records the device locations of completed parent outputs, and $\tau_t$ records device availability times. Thus $s_t$ summarizes the execution state that affects future switching, transfer, locality, and reuse.

Let $\mathcal{C}_t\subseteq V$ denote the set of completed stages at time $t$. The ready set is
\[
\mathcal{R}(s_t)=\{v\in V\setminus \mathcal{C}_t:\mathrm{Pa}(v)\subseteq \mathcal{C}_t\}.
\]
The feasible candidate set is
\[
\mathcal{F}(s_t)=\{(v,d): v\in \mathcal{R}(s_t),\ d\in A(v),\ \text{$(v,d)$ satisfies current resource feasibility}\}.
\]
We seek a policy $\pi$ that jointly determines stage placement, bounded shard execution, execution ordering, and planning decisions so as to minimize end-to-end workflow completion time:
\[
\min_{\pi}\; \mathbb{E}_{W\sim \mathcal{P}}\big[L(W;\pi)\big],
\]
where $L(W;\pi)$ denotes the completion time of workflow $W$ under policy $\pi$.

The key distinction is that execution cost is state-dependent rather than fixed. 
We therefore write the effective cost of placing stage $v$ on device $d$ under state $s_t$ as
\[
\hat{c}(v,d,s_t)
=
g\big(c_v(d),\, m(v),\, \phi(v),\, \mathrm{Pa}(v),\, \mathrm{Ch}(v),\, s_t\big),
\]
where $g(\cdot)$ captures how the current execution state modifies the nominal stage-device runtime through model residency, parent-output locality, prefix-related reuse, transfer cost, and bounded shard execution effects. 
The exact decomposition of this state-dependent cost is method-specific; \textsc{FATE} instantiates it with the planning scores, scheduling scores, and correction terms described in Section~\ref{sec:Method}. Thus, the problem is not merely heterogeneous but state-dependent: a decision at time $t$ changes the future execution state and thereby changes downstream switching cost, transfer cost, locality, and reuse opportunities.

\section{Method}
\label{sec:Method}
Our method, FATE, is built on a simple principle: a scheduling decision should optimize not only immediate execution efficiency, but also the quality of the future execution state that it induces. This view unifies several effects that are often treated as separate heuristics, including model locality, parent-child colocation, prefix-related reuse, and transfer minimization.

\subsection{Overview}

\textsc{FATE} consists of three components: (i) a CP-SAT-backed frontier planner~\cite{perron2023cp} whose candidate scores incorporate a limited downstream horizon; (ii) an incremental commit-and-advance planning procedure that moves the horizon forward level by level during schedule construction; and (iii) a state-conditional cost and scoring model that adjusts execution estimates according to the current system state.

For each planning step, \textsc{FATE} evaluates the current ready frontier using horizon-aware candidate scores derived from a limited downstream level view. It then invokes a CP-SAT planner to produce candidate stage placements and bounded multi-device execution decisions under the current state. The selected placements are materialized as executable stage-device assignments, and the runtime executor issues dependency-ready tasks according to the constructed per-device execution order. Rather than solving the full state-dependent scheduling problem exactly, \textsc{FATE} operationalizes future-state preservation through a practical decomposition into frontier planning, readiness-based execution, and state-conditional cost correction.

\subsection{Future-State-Aware Decomposition}

The main design principle of \textsc{FATE} is that a scheduling action should be evaluated not only by its immediate execution cost, but also by the quality of the future execution state that it induces. At a conceptual level, this tradeoff can be written as

\[
Q(s_t,a_t)
=
- C_{\mathrm{imm}}(s_t,a_t)
+
\gamma\,V_{\mathrm{future}}(s_{t+1}),
\]

where $C_{\mathrm{imm}}(s_t,a_t)$ denotes immediate execution cost and $V_{\mathrm{future}}(s_{t+1})$ summarizes the downstream value of the resulting execution state.

\textsc{FATE} does not learn or directly optimize this value function. Instead, it uses this decomposition to structure a practical approximation: a stage-wise planner whose candidate scores include downstream tail estimates over a limited level horizon, together with scheduling scores and cost-correction terms that account for model residency, dependency-aligned placement, prefix-related reuse, and bounded shard parallelism.

\subsection{Horizon-Aware Frontier Planning}

To reason about future-state tradeoffs explicitly, \textsc{FATE} uses a stage-wise CP-SAT planner. 
Planning is indexed by downstream DAG levels, but the solver itself is invoked on the current ready frontier rather than on all future levels simultaneously. 
% A limited downstream horizon $H$ influences current placement decisions through horizon-aware candidate scoring instead of introducing every future stage as an explicit solver variable.
We use a bounded horizon $H$ in the candidate scoring step. In the implementation, $H=1$ scores the current ready frontier, while larger values add discounted estimates from up to $H-1$ downstream DAG levels. Rather than expanding those future stages into explicit CP-SAT variables, \textsc{FATE} folds their estimated demand and reuse effects into horizon-aware scores attached to current-frontier candidates.

At planning step $j$, let $\mathcal{R}_j$ denote the current ready frontier. 
For each ready stage $v\in\mathcal{R}_j$, shard slot $k\in\{0,\dots,R(v)-1\}$, and eligible device $d\in A(v)$, the planner introduces a binary assignment variable $y_{v,k,d}$ indicating whether shard slot $k$ of stage $v$ is placed on device $d$. 
The frontier planner solves
\[
\max_y
\sum_{v\in\mathcal{R}_j}
\sum_{k=0}^{R(v)-1}
\sum_{d\in A(v)}
y_{v,k,d}\,\Psi(v,k,d\mid s_t,H),
\]
where $\Psi(v,k,d\mid s_t,H)$ is a state-aware placement score combining immediate execution quality with limited-horizon downstream tail estimates.

The planner enforces eligibility, device-capacity, and bounded-shard constraints. 
In particular, each device can receive at most one frontier assignment at a planning step, shard slot $0$ acts as the primary assignment for a stage, and higher-index shard slots can be enabled only after lower-index slots are enabled. 
These constraints allow \textsc{FATE} to model bounded multi-device execution without allowing unconstrained replication or speculative racing. Bounded shard execution partitions a stage's query batch across selected devices; it does not split a single query across devices.

After solving on the current frontier, \textsc{FATE} materializes the selected assignments as executable stage-device placements. As assigned stages complete, the execution state is updated; when the current frontier has been exhausted, the planner recomputes the next ready frontier and repeats.

Thus, the planner is rolling and stage-wise: future levels influence current choices through horizon-aware scores, but runtime does not solve one monolithic optimization problem over the entire DAG. 
Full decision variables, constraints, and implementation details are provided in Appendix~\ref{Technical_Appendix}.

\subsection{State-Aware Scoring}

The planner determines shard-slot placements and materializes them as executable stage-device assignments. To compare feasible assignments under the current execution state, \textsc{FATE} uses a future-state-aware score.

For a candidate stage-device assignment $(v,d)$, \textsc{FATE} computes
\[
\begin{aligned}
S(v,d \mid s_t)
={}& -\lambda_q \, C_{\text{wait}}(v,d,s_t)
-\lambda_s \, C_{\text{switch}}(v,d,s_t) -\lambda_{\mathrm{tr}} \, C_{\text{transfer}}(v,d,s_t) \\
&
+\lambda_c \, B_{\text{colo}}(v,d,s_t) +\lambda_p \, B_{\text{prefix}}(v,d,s_t)
+\lambda_r \, B_{\text{parallel}}(v,d,s_t).
\end{aligned}
\]
The terms capture waiting cost, model-switch cost, cross-device transfer cost, parent-child colocation benefit, prefix-related reuse benefit, and bounded multi-device shard-execution benefit.

\textsc{FATE} uses this score to rank feasible stage-device candidates during schedule construction:
\[
(v^*, d^*)=\arg\max_{(v,d)\in \mathcal{F}(s_t)} S(v,d\mid s_t),
\]
where $\mathcal{F}(s_t)$ denotes the feasible candidate set defined in Section~\ref{sec:formulation}. In implementation, selected assignments are materialized into per-device execution queues, and the runtime executor issues dependency-ready stages when their assigned devices become available. Detailed definitions of individual score terms are provided in the technical appendix.

\subsection{State-Conditional Cost Estimation}

The score terms used by the planner and readiness-based scheduling procedure are computed from a shared state-conditional cost estimator. Given a stage-device pair $(v,d)$ and state $s_t$, the estimator starts from the nominal runtime profile $c_v(d)$ and applies corrections for model residency, parent-output placement, prefix metadata, queue pressure, and bounded shard execution. These corrected estimates instantiate the cost and benefit terms in both the planner score $\Psi(v,k,d\mid s_t,H)$ and the scheduling score $S(v,d\mid s_t)$.

This separation is important: the planner uses the estimator to compare frontier assignments under limited downstream lookahead, while the executor uses the resulting stage-device order subject to runtime dependency readiness. Thus, the estimator is not a third scheduling objective; it is the common measurement layer that supplies state-dependent costs to both planning and schedule construction. 

Algorithm~\ref{alg:fate_main} summarizes the overall scheduling procedure. Expanded decision variables, constraints, scheduling-score term definitions, and procedural details are provided in Appendix~\ref{Technical_Appendix}.

\begin{algorithm}[t]
\caption{\textsc{FATE}: Future-State-Aware Scheduling}
\label{alg:fate_main}
\small
\begin{algorithmic}[1]
\State Initialize execution state $s_0$ and per-device execution queues
\While{there remain unassigned stages among active workflow instances}
    \State Construct current ready frontier $\mathcal{R}_j$
    \State Compute frontier candidate scores $\Psi(v,k,d\mid s_t,H)$
    \State Solve frontier CP-SAT planner on $(s_t,\mathcal{R}_j)$
    \State Materialize selected shard-slot placements into per-device execution queues
    \State Advance the planning frontier and update planning state
\EndWhile
\While{there remain unfinished executable stages}
    \State Issue dependency-ready queued stages when their assigned devices are available
    \State Update execution state $s_t$ as stages complete
\EndWhile

\end{algorithmic}
\end{algorithm}

\section{Experiments}
In this section, we present a comprehensive experimental evaluation to answer the following questions. First, does \textsc{FATE} improve end-to-end workflow performance over strong practical heuristics, classical DAG-scheduling, and proxy adaptations of recent workflow-serving policies? Second, are these gains consistent across structurally different workflow groups and controlled prefix-reuse settings? Third, what mechanism-level changes (e.g., model switching, cross-device dependencies, and prefix/locality proxies) explain the observed improvements?

\subsection{Experimental Setup}

\paragraph{Benchmarks.} We evaluate \textsc{FATE} on two benchmark families. First, we derive a  WfCommons/ WfInstances-based benchmark~\cite{coleman2022wfcommons}. Because there is not yet a widely adopted benchmark tailored to heterogeneous multi-LLM workflow scheduling, we construct a structured evaluation substrate by lifting workflow DAGs into LLM-stage execution graphs. The lifting preserves dependency structure among the lifted stage groups and augments each stage with a predefined role template, model assignment, and runtime proxy profile for execution, caching, and cross-device transfer. The templates are fixed before evaluation and applied consistently across methods. Appendix~\ref{app:benchmark} provides the lifting procedure, stage-role templates, model-assignment rules, runtime-proxy construction, and prefix-reuse generation details. Second, we construct a controlled prefix-reuse benchmark by pairing workflow-style DAG templates with controlled long-context workloads whose shared-prefix repeat ratios are set to $0$, $0.25$, $0.5$, and $1.0$.

Following Halo's batch-query workflow-serving setting~\cite{shen2025batch}, each benchmark item is a workflow template evaluated on a batch of independent query instances. We evaluate each workflow template under two batch-size configurations, 16 and 32 independent query instances, to test robustness under heavier batched workflow concurrency.

\paragraph{Compared methods.}
We compare \textsc{FATE} against four baseline types under the same runtime framework, hardware setting, and workflow templates.
Our primary workflow-serving baseline is \textbf{Halo}~\cite{shen2025batch}, implemented with Halo's original search-based scheduling algorithm.
We also include \textbf{RoundRobin} as a simple dispatch baseline, \textbf{KVFlow}~\cite{pankvflow} as a future-reuse-aware cache-oriented policy, \textbf{Helix}~\cite{mei2025helix} as a heterogeneity-aware placement policy, and \textbf{HEFT}~\cite{topcuoglu2002performance} as a classical DAG-scheduling baseline.
All methods are implemented in our unified workflow-serving evaluation framework under the same execution environment.

\paragraph{Implementation summary.}
\textsc{FATE} uses stage-wise CP-SAT planning over downstream DAG levels, horizon-aware frontier commit-and-advance, explicit state-aware score terms for waiting, switching, transfer, colocation, prefix reuse, and bounded shard parallelism, and refined runtime cost modeling for model residency, switch overhead, and parent readiness. In deep heterogeneous workflows, the implementation further uses model-specialized placement preferences; in cache-dominant same-model settings, it enables bounded multi-device shard execution when beneficial.

\paragraph{Metrics.}
Our primary end-to-end metrics are normalized makespan and normalized P95 query completion latency . Both are computed per workflow relative to RoundRobin and aggregated across instances using the geometric mean. For mechanism diagnostics, we report three per-task proxy metrics: X-Dev Edge, Cache Score, and Model Cont. X-Dev Edge is the number of cross-device parent edges normalized by the total number of workflow tasks, Cache Score is the estimated number of prefix-cache hits normalized by the total number of workflow tasks, and Model Cont. is the number of same-model continuations normalized by the total number of workflow tasks. These are diagnostic proxies rather than end-to-end objectives: X-Dev Edge approximates transfer pressure, Cache Score approximates prefix/cache affinity, and Model Cont. approximates model-residency continuity. 
Formal metric definitions are provided in Appendix~\ref{app:metrics}.

\subsection{Main Results}

Table~\ref{tab:overall_real_dag} reports the main results on the workflow-DAG benchmark. \textsc{FATE} achieves the best end-to-end performance across both normalized makespan and normalized P95 latency. Its normalized makespan is $0.675$, compared with $0.741$ for Helix, $0.748$ for KVFlow, $0.791$ for HEFT, $0.902$ for Halo, and $1.000$ for RoundRobin. The same pattern holds for tail performance: \textsc{FATE} obtains a normalized P95 latency of $0.677$, improving over Helix at $0.742$ and KVFlow at $0.750$.

The mechanism-level diagnostics help explain these gains. \textsc{FATE} achieves the lowest X-Dev Edge score, indicating fewer cross-device parent dependencies, and the highest Cache Score and Model Cont. values, indicating stronger prefix/cache affinity and model-residency continuity. These trends match the intended behavior of future-state-aware scheduling: preserve useful execution state when possible, avoid avoidable cross-device fragmentation, and maintain downstream reuse opportunities rather than optimizing only the immediate step.

\begin{figure}[t]
    \centering
    \includegraphics[width=0.62\linewidth]{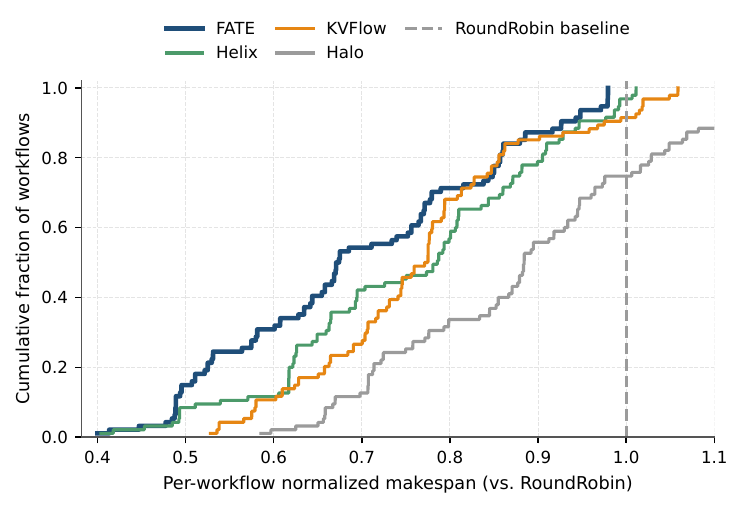}
    \caption{ECDF of per-workflow normalized makespan on the workflow-DAG benchmark. The x-axis is makespan normalized by RoundRobin for each workflow, and the y-axis is the cumulative fraction of workflows. Curves farther to the left indicate better performance.}
    \label{fig:real_dag_ecdf}
\end{figure}

Figure~\ref{fig:real_dag_ecdf} shows the distribution of per-workflow normalized makespan against RoundRobin. The ECDF compares \textsc{FATE} with KVFlow, Helix, Halo, and RoundRobin. The distributional view confirms that the aggregate improvement is not driven only by a few favorable cases; \textsc{FATE} shifts a broad fraction of workflows toward lower normalized makespan.

The overhead of planning is small in practice. In our runs, every CP-SAT solve finishes within $0.08$s under the evaluated workloads, which is small relative to end-to-end workflow completion time. This suggests that the observed gains do not come at a prohibitive scheduling cost.

\begin{table}[t]
\centering
\small
\caption{Overall results on the workflow-DAG benchmark. Normalized makespan and normalized P95 query completion latency are computed per workflow relative to RoundRobin and aggregated using the geometric mean. X-Dev Edge, Cache Score, and Model Cont. are per-task mechanism proxies; formal definitions are provided in Appendix~\ref{app:metrics}. Lower is better for normalized metrics and X-Dev Edge; higher is better for Cache Score and Model Cont.}

\label{tab:overall_real_dag}
\begin{tabular}{lccccc}
\toprule
& \multicolumn{2}{c}{\textbf{End-to-End}} & \multicolumn{3}{c}{\textbf{Mechanism}} \\
\cmidrule(lr){2-3} \cmidrule(lr){4-6}
Method & Norm. MS$\downarrow$ & Norm. P95$\downarrow$ & X-Device Edge$\downarrow$ & Cache Score$\uparrow$ & Model Cont.$\uparrow$ \\
\midrule
\textsc{FATE} (Ours)      & \textbf{0.675} & \textbf{0.677}  & \textbf{1.191} & \textbf{0.129} & \textbf{0.573} \\
KVFlow             & 0.748 & 0.750 & 1.417 & 0.124 & 0.492 \\
Helix              & 0.741 & 0.742 & 1.235 & 0.107 & 0.460 \\
Halo          & 0.902 & 0.907 & 1.242 & 0.081 & 0.374 \\
HEFT               & 0.791 & 0.791 & 1.433 & 0.073 & 0.449 \\
RoundRobin         & 1.0 & 1.0 & 1.557 & 0.075 & 0.332 \\
\bottomrule
\end{tabular}
\end{table}

\begin{table}[t]
\centering
\begin{minipage}{0.45\linewidth}
\centering
\small
\captionof{table}{Controlled prefix-reuse study. Values are geometric-mean makespan normalized by Halo at repeat ratio $0$. Lower is better.}
\label{tab:prefix_reuse_suite}
\begin{tabular}{lcccc}
\toprule
Method & 0 & 0.25 & 0.5 & 1.0 \\
\midrule
Halo          & 1.000 & 0.991 & 0.984 & 0.981 \\
KVFlow        & 0.702 & 0.695 & 0.693 & 0.690 \\
\textsc{FATE} & \textbf{0.596} & \textbf{0.594} & \textbf{0.593} & \textbf{0.587} \\
\bottomrule
\end{tabular}
\end{minipage}
\hfill
\begin{minipage}{0.5\linewidth}
\centering
\small
\captionof{table}{Ablation study on lifted workflow DAGs. Degradation is relative to full \textsc{FATE}.}
\label{tab:ablation}
\begin{tabular}{lcc}
\toprule
Variant & Norm. MS$\downarrow$ & Deg. \\
\midrule
Full \textsc{FATE}      & \textbf{0.675} & -- \\
w/o future planning     & 0.748 & +10.82\% \\
w/o locality terms      & 0.706 & +4.63\% \\
w/o same-model bonus    & 0.690 & +2.20\% \\
w/o prefix terms        & 0.685 & +1.41\% \\
w/o shard parallelism   & 0.687 & +1.82\% \\
\bottomrule
\end{tabular}
\end{minipage}
\end{table}

\subsection{Controlled Prefix-Reuse and Ablation Studies}

\paragraph{Controlled prefix reuse.}
Table~\ref{tab:prefix_reuse_suite} evaluates a controlled prefix-reuse benchmark in which the shared-prefix repeat ratio is directly manipulated. Unlike the workflow-DAG benchmark, this suite is designed primarily for behavioral analysis: it isolates whether increasing prefix reuse alone explains the performance gap. All values are normalized by Halo at repeat ratio $0$, so changes across columns reflect both scheduler differences and the effect of increasing shared-prefix availability.

\textsc{FATE} remains the strongest method across all repeat ratios, with normalized makespan staying between $0.587$ and $0.596$. KVFlow is competitive, as expected for a reuse-aware baseline, but it remains consistently behind \textsc{FATE}: the gap is $0.596$ versus $0.702$ at repeat ratio $0$ and $0.587$ versus $0.690$ at repeat ratio $1.0$. The small change across repeat ratios also shows that prefix reuse alone is not the dominant explanation for end-to-end workflow performance in this suite. \textsc{FATE}'s advantage comes from scheduling decisions that jointly preserve future model residency, dependency locality, and shard opportunities rather than from cache reuse alone.

\paragraph{Ablations.}
Table~\ref{tab:ablation} evaluates the contribution of the main components in \textsc{FATE}. The no-future-planning variant is a strong internal baseline: it retains the same state-aware cost terms but removes downstream lookahead. Its degradation from $0.675$ to $0.748$ is the largest ablation, suggesting that the gain is not merely from adding locality or cache-aware terms, but from using them in future-aware planning. Removing locality terms also degrades performance to $0.706$, supporting the importance of parent-child placement continuity and transfer-aware scheduling.

The remaining ablations show that prefix awareness, same-model preference, and bounded shard parallelism each contribute smaller but measurable gains. Removing prefix-aware terms increases normalized makespan to $0.685$, while removing shard parallelism increases it to $0.687$ and removing the same-model bonus increases it to $0.690$. These results are consistent with the broader thesis of the paper: no single mechanism fully explains the improvement. \textsc{FATE} works because it jointly preserves multiple dimensions of execution state.

\subsection{Takeaways}
The experiments support a more precise conclusion than a simple ``solver beats heuristics'' claim. On the WfCommons-derived workflow-DAG benchmark, \textsc{FATE} improves end-to-end performance, shifts the ECDF of per-workflow makespan toward lower latency, and degrades most when future-state planning is removed. Under controlled prefix reuse, it also remains stronger than reuse-aware scheduling alone. These results should be interpreted as evidence that workflow-state-aware scheduling is effective in our controlled lifted-DAG evaluation, rather than as a claim of full production-system dominance. The mechanism diagnostics and ablations suggest that the gain comes from jointly balancing model residency, dependency locality, transfer structure, prefix reuse, and bounded shard execution.

\section{Related Work}

\paragraph{LLM serving systems.}
Recent LLM serving systems improve efficiency through KV-cache management, structured execution, phase-aware scheduling, and disaggregation. Representative examples include vLLM, which introduced PagedAttention for efficient KV memory management~\cite{kwon2023efficient}; SGLang for structured program execution and reuse~\cite{zheng2024sglang}; Sarathi-Serve for chunked prefill scheduling~\cite{agrawal2024taming}; and DistServe and Mooncake for disaggregated or KVCache-centric serving architectures~\cite{zhong2024distserve,qin2024mooncake}. These systems substantially improve inference efficiency, but they primarily target single-request or request-level serving rather than heterogeneous multi-stage workflow DAGs.

\paragraph{Workflow-aware and agentic serving.}
More recent work has recognized that many LLM applications execute as dependent workflows rather than isolated calls~\cite{luo2025autellix,chaudhry2025murakkab,laju2026nalar}. Halo formulates such workloads as query-plan DAGs and optimizes execution at the workflow level~\cite{shen2025batch}. KVFlow studies workflow-aware prefix/KV management and focuses on preserving entries likely to be reused by future workflow steps~\cite{pankvflow}. Helix emphasizes heterogeneity-aware serving and placement~\cite{mei2025helix}. Our work is closest to this line, but differs in scope: rather than focusing on reuse or placement alone, we study execution-state-aware placement and scheduling under heterogeneous devices, state-dependent switching and transfer costs, and bounded multi-device shard execution.

\paragraph{Heterogeneous DAG scheduling.}
Classical DAG scheduling methods such as HEFT~\cite{topcuoglu2002performance} remain natural reference points because our workloads are structured DAGs executed on heterogeneous devices. However, unlike traditional task graphs with mostly static execution and communication costs, LLM workflow stages induce strongly state-dependent costs through model residency, parent-child locality, and prefix-related reuse. This makes static ranking-based scheduling insufficient in many regimes, motivating a scheduler that reasons explicitly about the future execution state created by current decisions.

\section{Limitations}

This work has several limitations. First, although we evaluate \textsc{FATE} on both workflow-DAG and controlled prefix-reuse benchmarks, the current study does not yet cover the full range of deployment conditions, such as bursty arrivals or stronger profile drift. More broadly, there is not yet a widely adopted benchmark tailored specifically to heterogeneous multi-LLM workflow scheduling. Our current evaluation therefore derives real-DAG workloads from WfCommons/WfInstances and lifts them into collapsed LLM-stage execution graphs with attached stage roles, model assignments, and runtime cost proxies, while controlled prefix-reuse structure is introduced separately for behavioral analysis. This provides a practical and structured evaluation substrate, but not yet a community-standard benchmark. Second, our KVFlow-, Helix-, and HEFT-style comparisons are policy adaptations implemented inside a common workflow-serving framework, not full-system reproductions of the original serving stacks. Third, some of our mechanism-level diagnostics are proxies rather than direct hardware counters: cross-device parent edges approximate transfer pressure rather than measuring actual transferred bytes, and the prefix/locality proxy approximates reuse behavior rather than measuring true KV-cache hit rates. These proxy metrics are used only for mechanism analysis; our primary conclusions are based on direct end-to-end metrics such as normalized makespan and workflow completion time.

\section{Conclusion}

This paper studies a question that is increasingly important in modern LLM systems: when applications execute as heterogeneous multi-stage workflows, what should the scheduler optimize? We argue that execution state — the evolving configuration of model residency, parent-child locality, reusable prefix structure, and constrained downstream reachability — is an important complement to instantaneous load and queue state in heterogeneous workflow scheduling.

We formalize execution-state-aware placement and scoring for heterogeneous workflow DAGs and instantiate it with \textsc{FATE}, a CP-SAT-backed scheduler with horizon-aware planning and state-conditional scoring. Empirically, \textsc{FATE} improves over strong practical heuristics, classical DAG scheduling, and proxy adaptations of recent workflow-serving policies on both workflow-DAG and controlled prefix-reuse benchmarks. The gains are accompanied by fewer cross-device parent edges, higher same-model continuation, and stronger prefix/locality proxies, indicating that the benefit comes from preserving useful future execution state rather than optimizing only the current step.

The main takeaway is that future-state preservation should be treated as an important optimization objective for heterogeneous LLM workflow serving. A scheduler that reasons jointly about switching, locality, and reuse can outperform baselines that optimize only one of these factors in isolation.

% \begin{ack}
% Use unnumbered first level headings for the acknowledgments. All acknowledgments

% Do {\bf not} include this section in the anonymized submission, only in the final paper. You can use the \texttt{ack} environment provided in the style file to automatically hide this section in the anonymized submission.
% \end{ack}

%\section*{Acknowledgments}
%The research is partially supported by Quantum Science and Technology-National Science and Technology Major Project (QNMP) 2021ZD0302900 and China National Natural Science Foundation with No. 92567301, 62132018, 62231015, "Pioneer" and "Leading Goose" R\&D Program of Zhejiang, 2023C01029, and 2023C01143.

\bibliographystyle{unsrtnat}
\bibliography{references}

@inproceedings{kwon2023efficient,
  title={Efficient memory management for large language model serving with pagedattention},
  author={Kwon, Woosuk and Li, Zhuohan and Zhuang, Siyuan and Sheng, Ying and Zheng, Lianmin and Yu, Cody Hao and Gonzalez, Joseph and Zhang, Hao and Stoica, Ion},
  booktitle={Proceedings of the 29th symposium on operating systems principles},
  pages={611--626},
  year={2023}
}

@article{zheng2024sglang,
  title={Sglang: Efficient execution of structured language model programs},
  author={Zheng, Lianmin and Yin, Liangsheng and Xie, Zhiqiang and Sun, Chuyue and Huang, Jeff and Yu, Cody H and Cao, Shiyi and Kozyrakis, Christos and Stoica, Ion and Gonzalez, Joseph E and others},
  journal={Advances in neural information processing systems},
  volume={37},
  pages={62557--62583},
  year={2024}
}

@inproceedings{agrawal2024taming,
  title={Taming Throughput-Latency tradeoff in LLM inference with Sarathi-Serve},
  author={Agrawal, Amey and Kedia, Nitin and Panwar, Ashish and Mohan, Jayashree and Kwatra, Nipun and Gulavani, Bhargav and Tumanov, Alexey and Ramjee, Ramachandran},
  booktitle={18th USENIX symposium on operating systems design and implementation (OSDI 24)},
  pages={117--134},
  year={2024}
}

@inproceedings{zhong2024distserve,
  title={DistServe: Disaggregating prefill and decoding for goodput-optimized large language model serving},
  author={Zhong, Yinmin and Liu, Shengyu and Chen, Junda and Hu, Jianbo and Zhu, Yibo and Liu, Xuanzhe and Jin, Xin and Zhang, Hao},
  booktitle={18th USENIX Symposium on Operating Systems Design and Implementation (OSDI 24)},
  pages={193--210},
  year={2024}
}

@inproceedings{mei2025helix,
  title={Helix: Serving large language models over heterogeneous gpus and network via max-flow},
  author={Mei, Yixuan and Zhuang, Yonghao and Miao, Xupeng and Yang, Juncheng and Jia, Zhihao and Vinayak, Rashmi},
  booktitle={Proceedings of the 30th ACM International Conference on Architectural Support for Programming Languages and Operating Systems, Volume 1},
  pages={586--602},
  year={2025}
}

@article{topcuoglu2002performance,
  title={Performance-effective and low-complexity task scheduling for heterogeneous computing},
  author={Topcuoglu, Haluk and Hariri, Salim and Wu, Min-You},
  journal={IEEE transactions on parallel and distributed systems},
  volume={13},
  number={3},
  pages={260--274},
  year={2002},
  publisher={IEEE}
}

@inproceedings{perron2023cp,
  title={The CP-SAT-LP solver (invited talk)},
  author={Perron, Laurent and Didier, Fr{\'e}d{\'e}ric and Gay, Steven},
  booktitle={29th International Conference on Principles and Practice of Constraint Programming (CP 2023)},
  pages={3--1},
  year={2023},
  organization={Schloss Dagstuhl--Leibniz-Zentrum f{\"u}r Informatik}
}

@article{luo2025autellix,
  title={Autellix: An efficient serving engine for llm agents as general programs},
  author={Luo, Michael and Shi, Xiaoxiang and Cai, Colin and Zhang, Tianjun and Wong, Justin and Wang, Yichuan and Wang, Chi and Huang, Yanping and Chen, Zhifeng and Gonzalez, Joseph E and others},
  journal={arXiv preprint arXiv:2502.13965},
  year={2025}
}

@article{chaudhry2025murakkab,
  title={Murakkab: Resource-efficient agentic workflow orchestration in cloud platforms},
  author={Chaudhry, Gohar Irfan and Choukse, Esha and Qiu, Haoran and Goiri, {\'I}{\~n}igo and Fonseca, Rodrigo and Belay, Adam and Bianchini, Ricardo},
  journal={arXiv preprint arXiv:2508.18298},
  year={2025}
}

@inproceedings{pankvflow,
  title={KVFlow: Efficient Prefix Caching for Accelerating LLM-Based Multi-Agent Workflows},
  author={Pan, Zaifeng and PATEL, AJJKUMAR and Shen, Yipeng and Hu, Zhengding and Guan, Yue and Li, Wan-Lu and Qin, Lianhui and Wang, Yida and Ding, Yufei},
  booktitle={The Thirty-ninth Annual Conference on Neural Information Processing Systems}
}

@article{laju2026nalar,
  title={Nalar: An agent serving framework},
  author={Laju, Marco and Son, Donghyun and Agarwal, Saurabh and Kedia, Nitin and Lee, Myungjin and Srinivasa, Jayanth and Akella, Aditya},
  journal={arXiv preprint arXiv:2601.05109},
  year={2026}
}

@article{shen2025batch,
  title={Batch Query Processing and Optimization for Agentic Workflows},
  author={Shen, Junyi and Wadlom, Noppanat and Lu, Yao},
  journal={arXiv preprint arXiv:2509.02121},
  year={2025}
}

@article{qin2024mooncake,
  title={Mooncake: A kvcache-centric disaggregated architecture for llm serving},
  author={Qin, Ruoyu and Li, Zheming and He, Weiran and Cui, Jialei and Tang, Heyi and Ren, Feng and Ma, Teng and Cai, Shangming and Zhang, Yineng and Zhang, Mingxing and others},
  journal={ACM Transactions on Storage},
  publisher={ACM New York, NY}
}

@article{coleman2022wfcommons,
  title={WfCommons: A framework for enabling scientific workflow research and development},
  author={Coleman, Tain{\~a} and Casanova, Henri and Pottier, Lo{\"\i}c and Kaushik, Manav and Deelman, Ewa and da Silva, Rafael Ferreira},
  journal={Future generation computer systems},
  volume={128},
  pages={16--27},
  year={2022},
  publisher={Elsevier}
}

@inproceedings{bai2024longbench,
  title={Longbench: A bilingual, multitask benchmark for long context understanding},
  author={Bai, Yushi and Lv, Xin and Zhang, Jiajie and Lyu, Hongchang and Tang, Jiankai and Huang, Zhidian and Du, Zhengxiao and Liu, Xiao and Zeng, Aohan and Hou, Lei and others},
  booktitle={Proceedings of the 62nd annual meeting of the association for computational linguistics (volume 1: Long papers)},
  pages={3119--3137},
  year={2024}
}

%%%%%%%%%%%%%%%%%%%%%%%%%%%%%%%%%%%%%%%%%%%%%%%%%%%%%%%%%%%%

\appendix

\section{Technical Appendix}\label{Technical_Appendix}
\FloatBarrier

Table~\ref{tab:notation_summary_1} summarizes the workflow and execution-state notation used below.
Table~\ref{tab:notation_summary_2} summarizes the planning, scoring, and cost notation used in the expanded method description.

\begin{table}[t]
\centering
\small
\setlength{\tabcolsep}{5pt}
\renewcommand{\arraystretch}{1.08}

\caption{Notation summary I: workflow and execution-state symbols.}
\label{tab:notation_summary_1}

\begin{tabular}{
>{\raggedright\arraybackslash}p{0.20\columnwidth}
>{\raggedright\arraybackslash}p{0.55\columnwidth}
}
\toprule
Symbol & Meaning \\
\midrule
$\mathcal{W}_t$ & Active workflow instances at time $t$. \\
$G=(V,E)$ & Workflow DAG with stages $V$ and precedence edges $E$. \\
$u,v$ & Workflow stages. \\
$(u,v)\in E$ & Dependency edge from parent $u$ to child $v$. \\
$m(v)$ & Model type used by stage $v$. \\
$A(v)\subseteq\mathcal{D}$ & Eligible device set of stage $v$. \\
$R(v)$ & Maximum bounded shard degree of stage $v$. \\
$c_v(d)$ & Base runtime of stage $v$ on device $d$. \\
$\phi(v)$ & Stage-local metadata and runtime annotations. \\
$\mathrm{Pa}(v),\mathrm{Ch}(v)$ & Parent and child stage sets of $v$. \\
$s_t=(\rho_t,\kappa_t,\ell_t,\tau_t)$ & Execution state at time $t$. \\
$\rho_t$ & Model residency state across devices. \\
$\kappa_t$ & Reusable prefix/cache-related state across devices. \\
$\ell_t(u)$ & Device holding the completed output of parent $u$. \\
$\tau_t$ & Device-availability component of the execution state. \\
$\tau_d^{\mathrm{free}}$ & Next available time of device $d$. \\
$\mathcal{C}_t$ & Completed stages by time $t$. \\
$\mathcal{R}(s_t)$ & Idealized ready set under state $s_t$. \\
$\mathcal{F}(s_t)$ & Feasible stage-device candidate set under state $s_t$. \\
\bottomrule
\end{tabular}
\end{table}

\begin{table}[t]
\centering
\small
\setlength{\tabcolsep}{5pt}
\renewcommand{\arraystretch}{1.2}

\caption{Notation summary II: planning, scoring, and cost symbols.}
\label{tab:notation_summary_2}

\begin{tabular}{
>{\raggedright\arraybackslash}p{0.23\columnwidth}
>{\raggedright\arraybackslash}p{0.72\columnwidth}
}
\toprule
Symbol & Meaning \\
\midrule
$\mathcal{R}_j$ & Current ready frontier at planning step $j$. \\
$H$ & Downstream lookahead horizon in DAG levels. \\
$k$ & Shard-slot index, with $k\in\{0,\ldots,R(v)-1\}$. \\
$y_{v,k,d}$ & Binary assignment variable for shard slot $k$ of stage $v$ on device $d$. \\
$\Psi(v,k,d\mid s_t,H)$ & Frontier-planning score under state $s_t$ and horizon $H$. \\
$S(v,d\mid s_t)$ & Runtime scheduling score for candidate assignment $(v,d)$. \\
$\mathcal{A}_{\mathrm{com}}$ & Committed action pool produced by the planner. \\
$\mathcal{Q}(s_t)$ & Dependency-ready queued action set derived from $\mathcal{A}_{\mathrm{com}}$. \\
$\hat{c}(v,d,s_t)$ & State-corrected cost estimate for placing stage $v$ on device $d$. \\
$C_{\mathrm{wait}}$ & Waiting cost induced by device unavailability. \\
$C_{\mathrm{switch}}$ & Model-switch cost. \\
$C_{\mathrm{transfer}}$ & Cross-device transfer cost from completed parents. \\
$B_{\mathrm{colo}}$ & Parent-colocation benefit. \\
$B_{\mathrm{prefix}}$ & Prefix/cache-related reuse benefit. \\
$B_{\mathrm{parallel}}$ & Benefit from bounded shard execution. \\
$\beta_{i,j}$ & Transfer coefficient from device $i$ to device $j$. \\
$\sigma(u,v)$ & Estimated intermediate-result size transferred from $u$ to $v$. \\
$\mathrm{grp}(v)$ & Shared-prefix group associated with stage $v$. \\
$\mathrm{Overlap}(\mathrm{grp}(v),d,s_t)$ & Estimated reusable shared-prefix overlap aligned with device $d$. \\
$\lambda_q,\lambda_s,\lambda_{\mathrm{tr}},\lambda_c,\lambda_p,\lambda_r$ & Weights on waiting, switching, transfer, colocation, prefix, and parallel terms. \\
\makecell[l]{
$\Delta_{\mathrm{switch}},\Delta_{\mathrm{transfer}},$ \\
$\Delta_{\mathrm{locality}},\Delta_{\mathrm{prefix}},$ \\
$\Delta_{\mathrm{parallel}}$
}
& State-dependent correction terms in the cost estimator. \\
\bottomrule
\end{tabular}
\end{table}

\subsection{Expanded Problem Formulation}

We use the same notation as in the main text. We consider an online system serving a dynamic set of workflow instances. Let $\mathcal{W}_t$ denote the set of active workflows at time $t$. For notational simplicity, stage- and workflow-level quantities are defined for a generic workflow instance, and workflow indices are suppressed when unambiguous. Runtime ready sets, frontier sets, and feasible action sets are understood as unions of the corresponding per-workflow sets over the active instances in $\mathcal{W}_t$.

Each workflow instance is represented as a directed acyclic graph (DAG) $G=(V,E)$, where each node $v\in V$ denotes a workflow stage and each directed edge $(u,v)\in E$ denotes a precedence dependency: stage $v$ may begin only after the output of stage $u$ is available. Each stage $v$ is characterized by
\[
\big(m(v),\, A(v),\, R(v),\, c_v,\, \phi(v),\, \mathrm{Pa}(v),\, \mathrm{Ch}(v)\big).
\]
Here, $m(v)$ denotes the model type, $A(v)\subseteq\mathcal{D}$ is the eligible device set, $R(v)\in\mathbb{Z}_{\ge 1}$ is the maximum bounded shard degree, and $c_v:A(v)\rightarrow\mathbb{R}_{+}$ denotes the base runtime profile across eligible devices, with
\[
c_{\mathrm{base}}(v,d)=c_v(d), \qquad d\in A(v).
\]
The feature vector $\phi(v)$ contains stage-local metadata such as prompt-side annotations, shared-prefix-group information, cache-related flags, and other runtime-side annotations. The sets $\mathrm{Pa}(v)$ and $\mathrm{Ch}(v)$ denote the parent and child stages of $v$, respectively.

The execution state at time $t$ is written as
\[
s_t=(\rho_t,\kappa_t,\ell_t,\tau_t),
\]
where $\rho_t$ records model residency on each device, $\kappa_t$ records reusable prefix-related metadata on each device, $\ell_t$ records the device locations of completed parent outputs, and $\tau_t$ records device availability times. Thus, $s_t$ summarizes the system state that shapes future switching, transfer, locality, and reuse.

Let $\mathcal{C}_t$ denote the set of completed stages by time $t$. In implementation, runtime also tracks stages that are currently executing and stages that have already been committed by the planner but have not yet completed. To keep the main-text formulation compact, these implementation-level sets are suppressed there. The idealized ready set in the main text is
\[
\mathcal{R}(s_t)
=
\{v\in V\setminus\mathcal{C}_t:\mathrm{Pa}(v)\subseteq \mathcal{C}_t\},
\]
while the implementation-level ready frontier further excludes stages that are already running or already committed.

The feasible candidate set is
\[
\mathcal{F}(s_t)
=
\{(v,d): v\in \mathcal{R}(s_t),\ d\in A(v),\ \text{$(v,d)$ satisfies current resource feasibility}\}.
\]

The abstract state-dependent cost function $g(\cdot)$ in the main text can be instantiated as an additive penalty-benefit decomposition:
\[
\begin{aligned}
\hat{c}(v,d,s_t)
={}&
c_{\mathrm{base}}(v,d)
+
c_{\mathrm{switch}}(v,d,s_t)
+
c_{\mathrm{transfer}}(v,d,s_t) \\
&-
b_{\mathrm{locality}}(v,d,s_t)
-
b_{\mathrm{prefix}}(v,d,s_t)
-
b_{\mathrm{parallel}}(v,d,s_t).
\end{aligned}
\]
The key property is that scheduling decisions reshape the future execution state and thereby alter downstream switching cost, transfer cost, locality, and reuse opportunities.

\subsection{Expanded Description of \textsc{FATE}}

\textsc{FATE} consists of three interacting components: a frontier CP-SAT planner with horizon-aware candidate scoring, a commit-and-advance schedule-construction procedure, and a state-conditional cost and scoring model. The planner reasons over the current ready frontier, selects stage-device assignments, and materializes them into executable per-device queues. The runtime executor then issues dependency-ready queued stages when their assigned devices become available.

At each planning step $j$, \textsc{FATE} operates on the current ready frontier rather than solving a monolithic optimization problem over the full workflow DAG. Let $\mathcal{R}_j$ denote the current ready frontier. Let $H$ denote a downstream level horizon used only to estimate future tail value during candidate scoring. That is, $H$ influences the evaluation of current frontier assignments through downstream lookahead, but future levels are not introduced as explicit CP-SAT decision variables.

For each ready stage $v\in\mathcal{R}_j$, each shard slot
\[
k \in \{0,\dots,R(v)-1\},
\]
and each eligible device $d\in A(v)$, we introduce a binary decision variable
\[
y_{v,k,d}\in\{0,1\}.
\]
The variable $y_{v,k,d}=1$ indicates that shard slot $k$ of stage $v$ is placed on device $d$ at the current planning step.

The planner solves the frontier placement problem
\[
\max_{y}\;
\sum_{v\in\mathcal{R}_j}
\sum_{k=0}^{R(v)-1}
\sum_{d\in A(v)}
y_{v,k,d}\,\Psi(v,k,d\mid s_t,H),
\]
where $\Psi(v,k,d\mid s_t,H)$ is a state-aware placement score combining immediate execution quality with downstream tail-value estimates. In implementation, $\Psi$ incorporates base execution quality together with switching, transfer, locality, prefix-related reuse, shard-parallelism, and model-specialization effects.

The decision variables are subject to device-capacity, eligibility, and bounded-shard constraints. First, each device can receive at most one shard-slot assignment in a single frontier planning wave:
\[
\sum_{v\in\mathcal{R}_j}\sum_{k=0}^{R(v)-1} y_{v,k,d} \le 1,
\qquad \forall d\in\mathcal{D}.
\]
This constraint applies to a single frontier planning wave; additional assignments can be made after the frontier is advanced or the planner is reinvoked.

Second, each shard slot of a stage can be assigned to at most one eligible device:
\[
\sum_{d\in A(v)} y_{v,k,d} \le 1,
\qquad
\forall v\in\mathcal{R}_j,\;
k\in\{0,\dots,R(v)-1\}.
\]

Third, shard slots are enabled monotonically, so higher-index shard slots can be selected only if lower-index slots are also selected:
\[
\sum_{d\in A(v)} y_{v,k+1,d}
\le
\sum_{d\in A(v)} y_{v,k,d},
\qquad
\forall v\in\mathcal{R}_j,\;
k<R(v)-1.
\]

Finally, assignments are restricted to eligible devices by defining variables only for $d\in A(v)$; equivalently, one may impose
\[
y_{v,k,d}=0,
\qquad
\forall v\in\mathcal{R}_j,\;
k\in\{0,\dots,R(v)-1\},\;
d\notin A(v).
\]

Together, these constraints ensure that each device receives at most one stage-slot assignment in the current planning wave, that each shard slot is assigned to at most one device, that primary assignments are selected before additional shard slots, and that higher-index shard slots can be enabled only after lower-index shard slots have been enabled.

After solving on the current frontier, \textsc{FATE} materializes the selected frontier assignments into executable per-device queues. When the current frontier has been exhausted, the planner advances the workflow state, recomputes the next ready frontier, and repeats. Thus, the planner is rolling and stage-wise: future levels influence current choices only through horizon-aware scoring, while solver decisions remain local to the current frontier.

\paragraph{Bounded shard execution.}
In our implementation, bounded multi-device execution is realized as query-level shard execution at the sub-batch level. When multiple shard slots are selected for a ready stage, the current query set of that stage is partitioned into disjoint query shards and distributed across devices. Each device executes the same stage on a different subset of queries. No single query is split across devices, and no query is redundantly executed on multiple devices for racing or speculative completion.

Aggregation is correspondingly simple: each device returns outputs for the queries in its assigned shard, and these outputs are merged back into the workflow state by query identity. Hence, bounded shard execution changes how the current batch is partitioned across devices, rather than requiring a cross-device reduction for a single query. Only stages with $R(v)>1$ are shardable. In practice, $R(v)$ acts both as a capability constraint and as a policy cap: partitionable stages may allow bounded shard execution, whereas merge or finalization stages typically use $R(v)=1$.

\subsection{Expanded Scheduling Score Definitions}

The main paper defines the Scheduling score as
\[
\begin{aligned}
S(v,d \mid s_t)
={}&
-\lambda_q \, C_{\mathrm{wait}}(v,d,s_t)
-\lambda_s \, C_{\mathrm{switch}}(v,d,s_t)
-\lambda_{\mathrm{tr}} \, C_{\mathrm{transfer}}(v,d,s_t) \\
&+
\lambda_c \, B_{\mathrm{colo}}(v,d,s_t)
+
\lambda_p \, B_{\mathrm{prefix}}(v,d,s_t)
+
\lambda_r \, B_{\mathrm{parallel}}(v,d,s_t).
\end{aligned}
\]
We now define the individual components in more detail.

Let $t$ denote the current scheduling decision time, and let $\tau_d^{\mathrm{free}}$ denote the next available time of device $d$. The waiting cost is
\[
C_{\mathrm{wait}}(v,d,s_t)=\max\{0,\tau_d^{\mathrm{free}}-t\}.
\]

The model-switch cost is
\[
C_{\mathrm{switch}}(v,d,s_t)=
\begin{cases}
0, & \text{if } m(v) \text{ is already resident on } d,\\
\kappa_{\mathrm{switch}}(m(v),d), & \text{otherwise.}
\end{cases}
\]
This term penalizes placements that require loading or activating a model that is not currently resident on the selected device.

The cross-device transfer cost is estimated by
\[
C_{\mathrm{transfer}}(v,d,s_t)
=
\sum_{u\in \mathrm{Pa}(v)}
\mathbf{1}[\ell_t(u)\neq d]\cdot \beta_{\ell_t(u),d}\cdot \sigma(u,v),
\]
where $\ell_t(u)$ denotes the device location of the completed output of parent stage $u$, $\sigma(u,v)$ denotes the estimated intermediate-result size transferred from parent $u$ to stage $v$, and $\beta_{i,j}$ is a device-pair transfer coefficient.

To capture structural locality beyond immediate transfer cost, we define
\[
B_{\mathrm{colo}}(v,d,s_t)
=
\frac{1}{|\mathrm{Pa}(v)|}
\sum_{u\in \mathrm{Pa}(v)} \mathbf{1}[\ell_t(u)=d]
,
\]
with the convention that $B_{\mathrm{colo}}(v,d,s_t)=0$ when $\mathrm{Pa}(v)=\emptyset$. This term rewards placements that keep a stage close to its completed parents and therefore preserve dependency-aligned locality.

Prefix-related reuse is estimated by
\[
B_{\mathrm{prefix}}(v,d,s_t)
=
\kappa_{\mathrm{prefix}}\cdot \mathrm{Overlap}(\mathrm{grp}(v),d,s_t),
\]
where $\mathrm{grp}(v)$ denotes the shared-prefix group of stage $v$, and $\mathrm{Overlap}(g(v),d,s_t)$ estimates the degree of reusable shared-prefix structure already aligned with device $d$. This term serves as a proxy for potential reuse rather than as an oracle of exact future cache hits.

Finally, bounded shard-execution benefit is modeled as
\[
B_{\mathrm{parallel}}(v,d,s_t)=\Delta_{\mathrm{parallel}}(v,d,s_t),
\]
where $\Delta_{\mathrm{parallel}}(v,d,s_t)$ estimates the completion-time reduction enabled by feasible shard-aware bounded execution under the current load and dependency structure.

\subsection{Expanded State-Conditional Cost Model}

The planner and schedule-construction rule rely on state-corrected cost estimates. This estimator supplies the numerical cost inputs used by both the planner score $\Psi$ and the scheduling score $S$; it is not a separate scheduling objective. In expanded form, the estimator is written as
\[
\begin{aligned}
\hat{c}(v,d,s_t)
={}&
c_{\mathrm{base}}(v,d)
+
\Delta_{\mathrm{switch}}(v,d,s_t)
+
\Delta_{\mathrm{transfer}}(v,d,s_t) \\
&-
\Delta_{\mathrm{prefix}}(v,d,s_t)
-
\Delta_{\mathrm{locality}}(v,d,s_t)
-
\Delta_{\mathrm{parallel}}(v,d,s_t).
\end{aligned}
\]
The correction terms capture whether the required model is already resident, whether parent outputs are local or remote, whether reusable shared-prefix structure is aligned with the candidate device, whether the placement preserves downstream structural locality, and whether bounded shard execution reduces completion time or changes downstream contention.

This state-conditional view is essential to the formulation: two otherwise identical stage-device pairs may induce different downstream costs because they produce different future execution states. The planner uses these corrected estimates to compare frontier assignments under limited downstream lookahead, while the runtime executor follows the materialized stage-device order subject to dependency readiness and device availability.

\subsection{Full Procedural Description}

Algorithm~\ref{alg:fate_full} provides a more detailed procedural description than the compact version shown in the main paper. Selected placements are first materialized into a committed action pool $\mathcal{A}_{\mathrm{com}}$; from this pool, the runtime derives a dependency-ready queued action set $\mathcal{Q}(s_t)$, distinct from the feasible candidate set $\mathcal{F}(s_t)$ in the formulation, and issues queued stages once their dependencies and assigned devices are ready.

\begin{algorithm}[h]
\caption{\textsc{FATE}: Full Scheduling Procedure}
\label{alg:fate_full}
\small
\begin{algorithmic}[1]
\State Initialize execution state $s_0$
\State Initialize committed action pool $\mathcal{A}_{\mathrm{com}}\gets\emptyset$
\While{there remain unfinished stages among active workflow instances}
    \If{$\mathcal{A}_{\mathrm{com}}$ contains no feasible ready action}
        \State Construct current ready frontier $\mathcal{R}_j$
        \For{each $v\in\mathcal{R}_j$, $k\in\{0,\dots,R(v)-1\}$, $d\in A(v)$}
            \State Compute state-corrected cost $\hat{c}(v,d,s_t)$
            \State Compute horizon-aware candidate score $\Psi(v,k,d\mid s_t,H)$
        \EndFor
        \State Solve frontier CP-SAT planner on $(s_t,\mathcal{R}_j)$
        \State Obtain shard-slot placements on eligible devices
        \State Commit selected placements into $\mathcal{A}_{\mathrm{com}}$
    \EndIf

    \State Construct dependency-ready queued action set $\mathcal{Q}(s_t)$ from $\mathcal{A}_{\mathrm{com}}$
    \State Issue ready queued actions according to the materialized per-device order and device availability

    \State Update execution state as stages complete
    \State Remove completed or infeasible actions from $\mathcal{A}_{\mathrm{com}}$
\EndWhile
\end{algorithmic}
\end{algorithm}

\section{Baseline Adaptations}

To compare scheduling logic under a common execution substrate, we implement all baselines in the same serving stack. Our \textbf{Halo} baseline is a faithful reproduction of the original Halo scheduling policy in this stack. For other prior methods whose full systems are not directly compatible with our abstraction, we implement best-effort adaptations rather than full-system reproductions. This is necessary because, to our knowledge, there is no off-the-shelf system that jointly exposes our exact abstraction: heterogeneous multi-model workflow DAGs, prefix/cache-state proxies, model residency, parent-output locality, bounded shard execution, and a common executor. We therefore retain the core scheduling idea of each adapted prior method while fixing the surrounding workflow runtime, benchmark templates, model assignments, and execution environment.

\subsection{Classical and Practical Baselines}

We include the following classical and practical baselines. \textbf{Halo} serves as a faithful reproduction of the original workflow-serving heuristic and is implemented directly in our common runtime. \textbf{HEFT} is a classical heterogeneous DAG scheduling baseline adapted to our setting. \textbf{RR} performs round-robin placement and execution ordering.

\subsection{KVFlow-style Future-Reuse-Aware Cache Baseline}

Our KVFlow-style baseline is a best-effort adaptation that preserves the core idea of future-step-aware prefix priority and cache retention, combined with greedy scheduling. Concretely, it favors actions that preserve reusable prefix-related state and same-model affinity for likely near-future steps, but does not include the full execution-state-aware planner used by \textsc{FATE}.

\subsection{Helix-style Heterogeneity-Aware Placement Baseline}

Our Helix-style baseline is a best-effort adaptation that performs heterogeneity-aware earliest-finish placement with transfer- and switch-aware assignment. It explicitly accounts for device heterogeneity and immediate assignment cost, but does not model future workflow execution state beyond the current decision.

\subsection{Mapping from Original Methods to Implemented Baselines}

Table~\ref{tab:baseline_mapping} summarizes how our implemented baselines relate to the original methods.

\newcolumntype{L}[1]{>{\raggedright\arraybackslash}p{#1}}

\begin{table}[h]
\centering
\small
\setlength{\tabcolsep}{5pt}
\renewcommand{\arraystretch}{1.35}
\caption{Mapping from original methods to baselines implemented in our common serving stack. Halo is faithfully reproduced, while the other methods are adapted to the shared abstraction.}
\label{tab:baseline_mapping}
\begin{tabular}{L{2.4cm} L{5.0cm} L{4.1cm}}
\toprule
Original method & Core idea retained & Not modeled in adapted baseline \\
\midrule
KVFlow      
& Future-step-aware prefix priority, cache retention, same-model preference 
& Full KVFlow cache-management system and end-to-end system stack \\
\addlinespace[0.35em]

Helix       
& Heterogeneity-aware placement, earliest-finish assignment, transfer/switch-aware selection 
& Full heterogeneous serving system and explicit future workflow state modeling \\
\addlinespace[0.35em]

RoundRobin  
& Simple device rotation under dependency readiness 
& State-aware placement, prefix reuse, and heterogeneity-aware ranking \\
\addlinespace[0.35em]

HEFT        
& DAG ranking and earliest-finish assignment with heterogeneous compute and transfer proxies 
& LLM-specific prefix reuse and future execution-state preservation \\
\bottomrule
\end{tabular}
\end{table}

\subsection{Baseline Signal Access and Fairness Scope}

Table~\ref{tab:baseline_fairness_scope} summarizes the scheduling signals available
to each implemented policy. Our goal is not to claim that each adapted baseline is
a full reimplementation of the corresponding production system. Instead, all
methods share the same executor, workflow templates, model assignments, query
workloads, and hardware environment, and differ only in the scheduling signals and
lookahead mechanisms they are allowed to use. This setup isolates policy-level
behavior under a common runtime.

The main asymmetry is that several baselines already receive immediate state
signals such as residency or transfer cost, and the KVFlow-style baseline is
allowed to use prefix-reuse information. What they lack is \textsc{FATE}'s joint
use of horizon-aware frontier planning and runtime state preservation across
model residency, parent locality, prefix affinity, and bounded shard opportunities.
We therefore interpret the gains as evidence for joint future-state-aware
scheduling, rather than as evidence that every individual proxy signal is unique
to \textsc{FATE}.

\begin{table*}[t]
\centering
\small
\setlength{\tabcolsep}{4pt}
\renewcommand{\arraystretch}{1.18}
\caption{
Scheduling signals exposed to each implemented method.
All methods use the same workflow executor, generated templates, model assignments,
query workloads, and hardware environment.
``Residency,'' ``transfer,'' and ``prefix'' indicate whether the corresponding
proxy signal is available during scheduling.
}
\label{tab:baseline_fairness_scope}
\begin{tabularx}{\textwidth}{
>{\raggedright\arraybackslash}p{0.16\textwidth}
>{\centering\arraybackslash}p{0.12\textwidth}
>{\centering\arraybackslash}p{0.12\textwidth}
>{\centering\arraybackslash}p{0.10\textwidth}
>{\raggedright\arraybackslash}X
}
\toprule
Method & Residency & Transfer & Prefix & Lookahead mechanism \\
\midrule
\textsc{FATE}
& Yes
& Yes
& Yes
& Explicit future-state planning over the runtime frontier. \\

Halo-style
& Coarse
& No
& No
& Beam search over DAG assignments. \\

KVFlow-style
& Yes
& Partial
& Yes
& Cache/reuse-priority scheduling. \\

Helix-style
& Yes
& Yes
& No
& Heterogeneity-aware placement. \\

HEFT
& Yes
& Yes
& No
& Upward-rank DAG priority. \\

RoundRobin
& No
& No
& No
& No lookahead. \\
\bottomrule
\end{tabularx}
\end{table*}

\section{Additional Experimental Details}
\label{app:exp_details}

\subsection{Benchmark Construction Details}
\label{app:benchmark}

We construct two benchmark suites: a WfCommons/WfInstances-derived workflow-DAG suite and a controlled prefix-reuse suite. The WfCommons-derived suite uses WfCommons/WfInstances JSON workflow descriptions as a source of realistic dependency structure, while the controlled prefix-reuse suite keeps workflow templates fixed and varies the amount of repeated long-context prefix structure. Both suites are generated deterministically from fixed templates, stable hashes, and fixed construction seeds. The same generated workflow instances, model assignments, runtime proxies, and cache-related metadata are used for all compared scheduling methods.

\paragraph{WfCommons/WfInstances selection and filtering.}
For the WfCommons/WfInstances-based benchmark, we parse workflow instance JSON files from the configured raw-data root and construct directed acyclic graphs from their task dependency fields. The parser accepts the task, job, and node formats used across WfCommons releases, extracts task identifiers and parent dependencies, and applies explicit filtering by workflow family, track, collapsed graph size, graph depth, graph width, per-family caps, and total instance limits. The fixed paper manifest records the selected instance files and filter settings used for the reported results.

We use WfCommons/WfInstances as a source of realistic workflow dependency structure rather than as a direct trace of LLM execution. Raw workflow tasks are mapped into LLM-stage groups with attached model, runtime, communication, and cache-related metadata. This makes the benchmark suitable for controlled comparison of scheduling policies, while avoiding the stronger claim that each raw WfCommons task is replayed as a separate LLM invocation.

\paragraph{DAG lifting and stage-group construction.}
Each raw workflow instance is first parsed into a task DAG. To obtain an LLM-stage execution graph, the benchmark builder maps raw tasks into stage groups. When repeated raw task-name families occur, the builder deterministically collapses tasks with the same normalized task-name prefix into stage groups. The lifted DAG preserves induced dependencies between groups: an edge is added from group $u$ to group $v$ whenever any raw task in $u$ is a parent of any raw task in $v$. To avoid degenerate over-compression on very large workflows, the implementation caps the number of grouped stages when prefix collapse would otherwise produce too few groups.

Each lifted stage is then topologically annotated with indegree, outdegree, level, reverse depth, and level width. These structural annotations are used only to assign workflow roles and proxy costs. The resulting graph preserves dependency relationships among lifted stage groups, but it is not a one-to-one replay of every raw task.

\paragraph{Stage-role templates.}
The benchmark assigns each lifted stage a role template using deterministic structural rules. The builder uses information such as topological level, indegree, outdegree, fan-in/fan-out pattern, and normalized task-name family. Source-like stages and early wide stages are mapped to prompt-preparation, retrieval, routing, or decomposition roles; middle stages with substantial fan-out are mapped to worker-style roles; high-fanin stages are mapped to merge or aggregation roles; and sink-like or late stages are mapped to summarization, validation, verification, or final-synthesis roles.

The role assignment is not optimized for any scheduler. It is a fixed template used to attach LLM-serving attributes, including expected prefill/decode work, output-size proxy, communication weight, cache behavior, model candidates, and bounded data-parallel eligibility. All schedulers operate on the same role-annotated workflow instances.

\paragraph{Model assignment.}
Each lifted stage is assigned a model alias according to its role. Role-specific candidate sets introduce model heterogeneity across the workflow. The model aliases include Qwen-style, DeepSeek-style, and Llama-style 7--8B profiles. Some cache-oriented, long-context, or family-specific tracks use a fixed model alias to isolate locality and prefix-reuse behavior, while general stages choose from role-dependent candidate models using a deterministic stable hash. The WfCommons lifting script uses the fixed seed \texttt{20260423} for this assignment.

This deterministic assignment ensures that all scheduling methods see the same model distribution and that benchmark generation is reproducible. Model assignment is not optimized separately for any scheduler, and no scheduler is allowed to choose the model assigned to a stage.

\paragraph{Runtime, switch, transfer, and prefix/cache proxies.}
Because the source workflow traces are not LLM-serving traces, execution costs are represented through controlled proxies. Each model profile specifies model size, memory footprint, prefill/decode coefficients, and a model-switch penalty. Each stage role contributes a complexity multiplier, prefill/decode scaling, maximum-token proxy, output-size proxy, and communication weight.

Transfer pressure is modeled through parent-child dependencies and output-size proxies. A child stage placed on a different device from its completed parent incurs higher estimated transfer cost than a colocated child. Transfer cost is estimated from parent-output token proxies and a constant edge-transfer coefficient; it is therefore a relative locality proxy rather than a measured interconnect bandwidth model.

Model-residency continuity is modeled through switch-cost proxies and same-model continuation features. Prefix/cache behavior is represented by stage fields such as \texttt{shared\_prefix\_group}, \texttt{keep\_cache}, and \texttt{cache\_reuse}. The resulting cache score estimates whether a later stage can reuse a compatible prefix state on the same device; it is not a hardware KV-cache hit counter.

For each stage $v$, the benchmark also specifies a bounded shard degree $R(v)$ when stage-level parallel execution is applicable. Eligible stages may use up to $R(v)$ replicated execution slots, with $R(v) \leq 2$ in the reported experiments. This models limited stage-level data parallelism rather than tensor-parallel partitioning of a single model.

\paragraph{Controlled prefix-reuse workloads.}
In addition to the WfCommons-derived benchmark, we construct a controlled prefix-reuse suite from workflow-style DAG templates and LongBench-style~\cite{bai2024longbench} text pools. The generator creates repeated shared-prefix groups at target ratios $0$, $0.25$, $0.5$, and $1.0$, using a fixed prefix length and deterministic grouping. At higher repeat ratios, more queries are assigned to shared-prefix groups, creating more opportunities for cache- or prefix-aware policies to preserve useful state.

This suite is intended for behavioral analysis rather than as an additional real-trace benchmark. It isolates whether the advantage of \textsc{FATE} comes only from prefix reuse or from jointly balancing prefix affinity, parent-child locality, model-residency continuity, transfer pressure, and bounded shard execution. These workloads should therefore be interpreted as controlled stress tests for prefix reuse, not as additional WfCommons real-DAG traces.

\paragraph{Benchmark limitations.}
The benchmark preserves workflow DAG structure but does not claim to reproduce production LLM traces. Runtime, transfer, cache, and model-switch costs are proxy features derived from stage roles and model profiles. The benchmark also does not model bursty online arrivals, memory fragmentation, tokenizer-dependent KV-cache layout, or measured inter-device bandwidth. The purpose of the benchmark is therefore controlled policy comparison under fixed workflow and state proxies, rather than end-to-end prediction of serving latency on a particular cluster.

\subsection{Metrics and Aggregation}
\label{app:metrics}

Let $T_{m,i}$ denote the workflow makespan of method $m$ on workflow instance $i$, and let $T_{\mathrm{RR},i}$ denote the corresponding RoundRobin makespan. We report normalized makespan as the geometric mean
\[
\mathrm{NormMS}(m)
=
\exp\left(
\frac{1}{N}
\sum_{i=1}^{N}
\log
\frac{T_{m,i}}{T_{\mathrm{RR},i}}
\right).
\]

For query-level tail performance, let $L^{95}_{m,i}$ denote the p95 query completion latency of method $m$ on instance $i$. We report normalized p95 latency as
\[
\mathrm{NormP95}(m)
=
\exp\left(
\frac{1}{N}
\sum_{i=1}^{N}
\log
\frac{L^{95}_{m,i}}{L^{95}_{\mathrm{RR},i}}
\right).
\]

For mechanism diagnostics, we report three per-task proxy metrics. Cross-device edge rate is
\[
\mathrm{XDevEdge}(m)
=
\frac{
\sum_i \mathrm{cross\_device\_parent\_edges}_{m,i}
}{
\sum_i \mathrm{workflow\_tasks}_{i}
}.
\]
Cache score is
\[
\mathrm{CacheScore}(m)
=
\frac{
\sum_i \mathrm{prefix\_cache\_hits\_est}_{m,i}
}{
\sum_i \mathrm{workflow\_tasks}_{i}
}.
\]
Model continuation is
\[
\mathrm{ModelCont}(m)
=
\frac{
\sum_i \mathrm{same\_model\_continuations}_{m,i}
}{
\sum_i \mathrm{workflow\_tasks}_{i}
}.
\]

These mechanism metrics are used only to interpret scheduler behavior. Lower XDevEdge indicates fewer cross-device parent-child placements. Higher CacheScore indicates stronger estimated prefix/cache affinity. Higher ModelCont indicates stronger model-residency continuity.

\subsection{Ablation Definitions}

The ablation study disables one component of \textsc{FATE} at a time while keeping the benchmark, device setting, model assignment, and remaining scheduler parameters fixed.

The \emph{w/o future planning} variant disables the future-state planning path used by \textsc{FATE}: the effective planning horizon is collapsed to one, so the solver still assigns the current ready frontier but no longer uses downstream lookahead beyond the immediate frontier. It should therefore be interpreted as removing future-state planning rather than simply shortening a CP-SAT horizon. The \emph{w/o locality terms} variant disables parent-locality, transfer, and remote-parent placement terms. The \emph{w/o same-model bonus} variant removes the reward for continuing execution with a model already favored or resident on the selected device, while leaving model-switch cost enabled. The \emph{w/o prefix/cache terms} variant disables prefix/cache-aware scoring. The \emph{w/o shard parallelism} variant disables bounded multi-device shard execution.

These ablations are designed to test whether \textsc{FATE}'s gains come from a single dominant signal or from jointly preserving multiple dimensions of execution state.

\subsection{Evaluation Pipeline and Result Provenance}

The reported results are organized around an auditable pipeline. First, benchmark construction scripts produce fixed workflow templates and fixed experiment manifests. Second, the manifest runner executes every scheduler under the same runtime framework and exports one CSV file per experiment. Third, a single paper-table exporter consumes only those manifest-declared CSV files and computes the main table, ECDF inputs, ablation table, prefix-reuse table, and mechanism-metric summaries using the metric definitions in Appendix~\ref{app:metrics}. Results not reachable from the fixed manifest and exporter are not used for the paper tables.

This pipeline is intended to separate raw run logs from paper evidence. The manifest defines the set of workflow instances, scheduler variants, query counts, device configuration, and output CSV paths. The exporter defines the normalizers, geometric-mean aggregation, and mechanism proxy formulas. This makes the provenance of each reported number explicit: each table entry can be traced back to a method, workflow instance, raw CSV row, and aggregation formula.

\subsection{Reproducibility Details}

Benchmark construction uses deterministic rules and a fixed construction seed. The same generated workflow templates and query workloads are used for all compared methods. The main experiments are run on a single server with eight NVIDIA RTX 4090 GPUs, an AMD EPYC 7763 64-Core Processor, and 503 GB of system memory. A single benchmark run typically takes about 2--5 minutes, depending on the workflow suite, batch size, and scheduler configuration. The complete set of reported experiments takes approximately 2--3 days on this server. The released artifact README provides the exact environment setup, benchmark-generation commands, manifest paths, scheduler parameters, and exporter command used to reproduce the paper tables and figures from raw run logs.

\section{Additional Results}
\label{app:additional_results}

\subsection{Per-Family Breakdown}

Table~\ref{tab:family_breakdown} reports normalized makespan broken down by workflow family on the workflow-DAG benchmark. Values are geometric means within each family and are restricted to the strict intersection of completed runs across compared methods. This analysis should be interpreted over the lifted LLM-stage workflow templates used in the paper rather than over the raw WfCommons tasks.

The family-level breakdown reveals three broad structural regimes. First, cache-dominant families such as SeismicCrossCorrelation and Srasearch are shallow or small derived DAGs with low model diversity and strong same-model continuity. In these families, reuse-aware or locality-aware baselines already capture part of the exploitable structure, but the updated results show that cache-dominant structure does not eliminate the value of future-state-aware planning: \textsc{FATE} still achieves clear gains, especially on Srasearch, indicating that preserving downstream placement and model-state continuity can matter even when prefix reuse is abundant. Second, very wide branch-and-merge families such as BWA and BLAST expose relatively weak future-state continuity: many ready stages can be scheduled in parallel, cache-bearing structure is limited, and the main challenge is often immediate balancing or earliest-finish placement. In this regime, \textsc{FATE} is not uniformly superior, and strong non-\textsc{FATE} baselines can match or slightly exceed it. Third, deeper multi-model families such as Montage, SoyKB, and Nextflow create longer chains of downstream consequences in which current placement decisions affect later model residency, transfer, and merge locality over multiple stages. These remain favorable regimes for \textsc{FATE}, with some of the largest gains appearing on deep multi-model families and on structured cache-dominant families.

The remaining families, including 1000Genome and Cycles, lie between these extremes. They contain enough workflow structure for future-state-aware planning to help, but not enough to produce the largest margins seen in families such as Montage, SoyKB, or Srasearch. We therefore interpret the family-level results as supporting a workload-dependent claim: future-state preservation is most valuable when workflow structure creates nontrivial multi-step locality, residency, or reuse tradeoffs, and less decisive when the lifted DAG is highly symmetric and width-dominated. We also caution that the RNA-seq subset is small, so its family-level result should be treated as descriptive rather than statistically decisive.

We also note that the fixed paper manifests cap the lifted workflow size at 64 stages in order to keep the evaluation budget manageable. This appendix therefore characterizes the reported benchmark subset rather than the full range of possible workflow sizes. Larger lifted DAGs may expose longer downstream state dependencies and could change the relative value of future-state-aware planning, but this remains a question for future evaluation rather than a direct claim of the present study.

\begin{table}[h]
\centering
\small
\caption{Per-family normalized makespan on the workflow-DAG benchmark. Values are geometric means within each family, restricted to the strict intersection of completed runs across compared methods. Lower is better. ``Best non-\textsc{FATE} comparator'' denotes the strongest baseline within the same family.}

\label{tab:family_breakdown}
\begin{tabular}{lccc}
\toprule
Family & DAGs & \textsc{FATE} & Best non-\textsc{FATE} comparator \\
\midrule
1000Genome & 22 & 0.742 & 0.800 (Helix) \\
BLAST & 15 & 0.712 & 0.707 (HEFT) \\
BWA & 15 & 0.810 & 0.768 (KVFlow) \\
Cycles & 19 & 0.778 & 0.792 (KVFlow) \\
Montage & 12 & 0.535 & 0.600 (HEFT) \\
Nextflow & 9 & 0.708 & 0.725 (Helix) \\
RNA-seq & 3 & 0.688 & 0.738 (HEFT) \\
SeismicCrossCorrelation	& 11 & 0.762 & 0.847 (KVFlow) \\
SoyKB & 10 & 0.550 & 0.739 (HEFT) \\
Srasearch &	25 & 0.531 & 0.643 (Helix) \\
\bottomrule
\end{tabular}
\end{table}

\subsection{Controlled Conflict Stress Test}

As an appendix-only diagnostic extension, we construct a controlled conflict stress test to isolate a regime that the main controlled prefix-reuse suite does not emphasize. The main prefix-reuse suite is intentionally cache-friendly: same-model locality, parent locality, and repeated-prefix structure are mostly aligned. By contrast, the controlled conflict suite is designed so that these signals are intentionally put into tension.

Concretely, the workflow is a fixed chain of model-specialized stages that alternates across multiple model families while retaining cache-relevant state throughout the chain. The scheduler therefore repeatedly faces a multi-step tradeoff: placing the current stage near its immediate parent improves short-term locality, but may destroy the model residency that would be more valuable a few steps later when the workflow returns to an earlier model family. This is precisely the type of state tradeoff that a myopic locality heuristic handles poorly and that a limited-horizon future-state-aware scheduler is designed to address.

For artifact clarity, this suite is generated by the same workflow-benchmark generator used for the main controlled prefix-reuse study, but through a separate controlled-conflict template path. The artifact materializes four appendix-only templates, \texttt{workflow\_cache\_conflict\_000.yaml}, \texttt{workflow\_cache\_conflict\_025.yaml}, \texttt{workflow\_cache\_conflict\_050.yaml}, and \texttt{workflow\_cache\_conflict\_100.yaml}, corresponding to shared-prefix repeat ratios $0$, $0.25$, $0.5$, and $1.0$. These runs are evaluated under a separate appendix-only manifest and are not included in the main aggregate benchmark tables.

Table~\ref{tab:controlled_conflict} shows that this intentionally adversarial construction produces a much larger gap than the main controlled prefix-reuse suite. We emphasize that this is expected and is the reason the suite is useful: it amplifies the residency-versus-locality conflict rather than averaging it away across more benign workflows. The resulting gap should therefore be interpreted as mechanism-level supporting evidence, not as a headline benchmark intended to reflect natural workload frequencies.

The trend across repeat ratios is also informative. \textsc{FATE} already dominates at repeat ratio $0$, where additional shared-prefix reuse is absent, indicating that the main effect in this suite is not cache reuse alone. Increasing the repeat ratio from $0$ to $1.0$ yields only a modest further improvement for \textsc{FATE}, while the Halo and KVFlow-style baselines remain comparatively flat. This suggests that extra prefix reuse does not by itself resolve the induced conflict between immediate parent locality and preserving future model-specific execution state. Rather, the primary benefit comes from planning decisions that preserve useful downstream residency and placement structure over multiple future steps.

Because the construction is deliberately adversarial, we do not interpret the absolute gap in this table as representative of the main benchmark. Instead, we use it to support a narrower claim: when prefix locality, parent locality, and future model residency are intentionally placed in conflict, future-state-aware planning can matter much more than reuse-aware scheduling alone.

\begin{table}[h]
\centering
\small
\caption{Appendix-only controlled conflict stress test. The workflow intentionally alternates model-specialized stages along a chain, creating tension between immediate parent locality and future model residency. Values are normalized makespan relative to Halo within this suite. Lower is better.}
\label{tab:controlled_conflict}
\begin{tabular}{lcccc}
\toprule
Method & 0 & 0.25 & 0.5 & 1.0 \\
\midrule
Halo & 1.00 & 1.00 & 0.99 & 0.99 \\
KVFlow & 1.00 & 0.98 & 0.98 & 0.97 \\
\textsc{FATE} & 0.30 & 0.27 & 0.27 & 0.26 \\
\bottomrule
\end{tabular}
\end{table}

\subsection{Variance and Stability}

Benchmark construction, manifests, model assignment, and query workloads are fixed. We repeat all reported experiments twice under the same fixed manifests. The remaining run-to-run variation therefore comes mainly from serving-system initialization, model loading, GPU runtime effects, and process scheduling. For variance analysis, we use the strict intersection of workflow instances and scheduler variants for which both repeats completed.

Across the repeated experiments, the raw wall-clock makespan coefficient of variation has median $1.40\%$, mean $1.43\%$, and maximum $3.82\%$; the pairwise repeat spread has median $2.80\%$ and maximum $7.65\%$. After normalizing each repeated run by the RoundRobin run on the same workflow and repeat, the paired coefficient of variation has median $1.31\%$, mean $1.49\%$, and maximum $4.68\%$. We therefore treat large aggregate gaps as meaningful system-level effects, while reporting small differences near this noise level as descriptive rather than statistically decisive.

\subsection{Hyperparameter Sensitivity}
\label{app:hyperparameter-sensitivity}

We run a small hyperparameter sensitivity check to test whether \textsc{FATE}'s gains depend on a single fragile horizon or scoring weight. This experiment uses 30 family-balanced WfCommons-derived DAGs. The subset is selected deterministically from the workflow-DAG manifest by choosing instances close to a target size of 24 stages. This check is intended as a robustness analysis rather than as an additional tuning procedure.

Table~\ref{tab:hyperparameter_sensitivity} reports normalized makespan relative to RoundRobin on the same workflow. The default setting in the main experiments is $H=4$. The $H=0$ row disables future-state planning, while $H=1$, $H=2$, and $H=3$ reduce the lookahead horizon. The remaining rows perturb groups of runtime scoring terms: state terms scale same-model and model-switch terms, locality terms scale parent-colocation and transfer-related terms, and prefix terms scale prefix/cache reuse terms.

All tested configurations remain substantially below RoundRobin on this subset. The default $H=4$ setting is not uniquely optimal on these DAGs, which is expected because this subset is not used for tuning and because short horizons can be competitive on smaller or less deeply staged graphs. The important observation is that the performance stays in a narrow range under horizon and weight perturbations. We therefore interpret this experiment as evidence that \textsc{FATE}'s behavior is not explained by a single magic weight; the larger ablation suite remains the primary evidence for the contribution of each mechanism.

\begin{table}[t]
\centering
\small
\caption{Hyperparameter sensitivity on 30 WfCommons-derived DAGs. Values are geometric means of makespan normalized by RoundRobin on the same workflow. The last column divides each setting by the $H=4$ default. Lower is better.}
\label{tab:hyperparameter_sensitivity}
\begin{tabular}{lcc}
\toprule
Setting & Norm. MS$\downarrow$ & Rel. to $H=4$$\downarrow$ \\
\midrule
$H=0$ no future-state planning & 0.760 & 1.063 \\
$H=1$ & 0.706 & 0.987 \\
$H=2$ & 0.710 & 0.993 \\
$H=3$ & 0.726 & 1.015 \\
$H=4$ default & 0.715 & 1.000 \\
\midrule
State terms $\times 0.5$ & 0.697 & 0.975 \\
State terms $\times 1.5$ & 0.731 & 1.022 \\
Locality terms $\times 0.5$ & 0.723 & 1.011 \\
Locality terms $\times 1.5$ & 0.718 & 1.004 \\
Prefix terms $\times 0.5$ & 0.721 & 1.008 \\
Prefix terms $\times 1.5$ & 0.731 & 1.022 \\
\bottomrule
\end{tabular}
\end{table}

\subsection{Proxy-Cost Perturbation}
\label{app:proxy-perturbation}

We further test whether \textsc{FATE}'s advantage depends on a particular calibration of the runtime proxy profile. The benchmark templates use proxy costs for model switching, parent-transfer/communication pressure, and prefix/cache-reuse benefit. To stress these assumptions, we construct a 30-DAG family-stratified subset from the workflow-DAG benchmark and multiply each proxy class by $0.5$ and $2.0$ while keeping the workflow structure, query count, and scheduler configuration fixed.

Table~\ref{tab:proxy_perturbation} reports geometric-mean makespan normalized by RoundRobin under the same perturbation condition. \textsc{FATE} remains the strongest method across all perturbations. The absolute normalized makespan changes modestly, from $0.691$ under the default proxy profile to at most $0.718$ under the tested perturbations. This suggests that the observed advantage is not an artifact of a single switch-cost, transfer-cost, or prefix-benefit calibration. Instead, \textsc{FATE}'s joint treatment of future model residency, parent locality, transfer pressure, and prefix/cache affinity remains beneficial under both under-estimated and over-estimated proxy costs.

\begin{table}[t]
\centering
\small
\caption{Proxy-cost perturbation on a 30-DAG family-stratified subset of the benchmark. Values are geometric means of makespan normalized by RoundRobin under the same perturbation condition. Lower is better.}
\label{tab:proxy_perturbation}
\begin{tabular}{lccc}
\toprule
Condition & \textsc{FATE}$\downarrow$ & KVFlow$\downarrow$ & Helix$\downarrow$ \\
\midrule
Default & {0.691} & 0.749 & 0.773 \\
Switch cost $\times 0.5$ & {0.711} & 0.743 & 0.807 \\
Switch cost $\times 2.0$ & {0.718} & 0.755 & 0.800 \\
Transfer cost $\times 0.5$ & {0.715} & 0.755 & 0.804 \\
Transfer cost $\times 2.0$ & {0.710} & 0.737 & 0.779 \\
Prefix benefit $\times 0.5$ & {0.699} & 0.733 & 0.760 \\
Prefix benefit $\times 2.0$ & {0.695} & 0.739 & 0.775 \\
\bottomrule
\end{tabular}
\end{table}

The prefix rows should be interpreted as perturbations to the template-level prefix/cache benefit proxy, rather than as a replacement for the controlled prefix-reuse study in Table~\ref{tab:prefix_reuse_suite}. The controlled prefix-reuse suite varies the amount of reusable prefix structure, while this experiment varies the cost calibration attached to the existing workflow-DAG templates.

\subsection{Solver Overhead Breakdown}

Across the completed solver-backed workflow-DAG and ablation CSV files used for diagnostics, \textsc{FATE} records 2175 CP-SAT frontier solves. All recorded solves returned \texttt{OPTIMAL}. The mean solver wall time is $0.0059$s, median $0.0042$s, p95 $0.0142$s, and maximum $0.0754$s, with zero recorded objective gap. The overhead is reported separately from workflow runtime so that scheduling cost and execution latency are not conflated.

\begin{table}[h]
\centering
\small
\caption{CP-SAT frontier-planning overhead on completed solver-backed diagnostic runs.}
\label{tab:solver_overhead}
\begin{tabular}{lccccc}
\toprule
Solves & Optimal & Mean (s) & Median (s) & P95 (s) & Max (s) \\
\midrule
2175 & 2175 & 0.0059 & 0.0042 & 0.0142 & 0.0754 \\
\bottomrule
\end{tabular}
\end{table}

\subsection{Broader Impacts}

This work studies scheduling for heterogeneous LLM workflow serving systems. Its potential positive impacts are improved serving efficiency, lower resource waste, and reduced cost and energy use for workflow-based LLM applications. These improvements may make complex LLM workflows more accessible to researchers and smaller organizations with limited compute budgets.

The work does not introduce new model capabilities, new data collection mechanisms, or new user-facing applications. However, efficiency improvements may also increase total demand for LLM services, partially offsetting per-query energy savings. These risks are indirect and depend on the downstream application. Appropriate mitigations include applying the scheduler within systems that enforce existing safety policies, access controls, monitoring, rate limits, and privacy protections for the underlying LLM applications.

%%%%%%%%%%%%%%%%%%%%%%%%%%%%%%%%%%%%%%%%%%%%%%%%%%%%%%%%%%%%

% \newpage
% \input{checklist.tex}

\end{document}